\documentclass[12pt]{article}
\usepackage{amsmath}
\usepackage{graphicx,psfrag,epsf}
\usepackage{enumerate}
\usepackage{natbib}
\usepackage{url} % not crucial - just used below for the URL 
\usepackage{amsfonts}
\usepackage{amssymb}
\usepackage{makeidx}
\usepackage{multirow}
 \usepackage{rotating}
 \usepackage{algorithm2e}

\newcommand{\blind}{1}

\newcommand{\ex}[1]{\ensuremath{\mathbb{E}[#1]}}

\newcommand{\bd}[1]{\ensuremath{\mbox{\boldmath $#1$}}}

\DeclareMathOperator{\ND}{ND}
\DeclareMathOperator{\cont}{cont}
\DeclareMathOperator{\adjcont}{adjcont}

\begin{document}

\def\spacingset#1{\renewcommand{\baselinestretch}%
{#1}\small\normalsize} \spacingset{1}

%%%%%%%%%%%%%%%%%%%%%%%%%%%%%%%%%%%%%%%%%%%%%%%%%%%%%%%%%%%%%%%%%%%%%%%%%%%%%%

\if1\blind
{
  \title{\bf Improved inference for areal unit count data using graph-based optimisation}
  \author{Duncan Lee\\
    School of Mathematics and Statistics, University of Glasgow\\
    and \\
    Kitty Meeks\thanks{
    Both authors gratefully acknowledge funding from the Engineering and Physical Sciences Research Council (ESPRC) grant number EP/T004878/1 for this work, while the work of the second author was also funded by a Royal Society of Edinburgh Personal Research Fellowship (funded by the Scottish Government). The respiratory hospitalisation data were provided by Public Health Scotland. }\hspace{.2cm} \\
    School of Computing Science, University of Glasgow}
  \maketitle
} \fi

\if0\blind
{
  \bigskip
  \bigskip
  \bigskip
  \begin{center}
    {\LARGE\bf Improving inference for areal unit count data using graph-based optimisation}
\end{center}
  \medskip
} \fi

\bigskip
\begin{abstract}
Spatial correlation in areal unit count data is typically modelled by a set of random effects that are assigned a conditional autoregressive (CAR) prior distribution. The spatial correlation structure implied by this model depends on a binary neighbourhood matrix, where two random effects are assumed to be partially autocorrelated if their areal units share a common border, and are conditionally independent otherwise. This paper proposes a novel graph-based optimisation algorithm for estimating the neighbourhood matrix from the data,  by viewing the areal  units as the vertices of the graph and the neighbour relations as the set of edges. The superiority of our methodology compared to using the border sharing rule is comprehensively evidenced by simulation, before the method is applied to a new respiratory disease surveillance study in the Greater Glasgow and Clyde Health board in Scotland between 2011 and 2017.
\end{abstract}

\noindent%
\vfill

\newpage
\spacingset{1.45} % DON'T change the spacing!
\section{Introduction}
\label{sec:intro}

Spatio-temporal count data relating to a set of $K$ non-overlapping areal units for $N$ consecutive time periods are prevalent in many fields, including epidemiology (\citealp{stoner2019}) and social science (\citealp{bradley2016}). The spatial correlation in these data is typically modelled by conditional autoregressive (CAR, \citealp{besag1991}) models, which are specified as a prior distribution for a set of random effects  within a hierarchical model specification. A large volume of research has extended this class of models to the spatio-temporal domain, including  capturing: spatially correlated linear time trends (\citealp{Bernardinelli1995}); time period specific spatially correlated surfaces (\citealp{waller1997}); and a temporally evolving spatial surface  (\citealp{rushworth2014}).

The spatial autocorrelation structure implied by  these spatio-temporal CAR models depends on a $K\times K$ neighbourhood matrix  $\mathbf{W}$, which specifies which pairs of areal units are close together in space. A binary specification is typically adopted, where $w_{kj}=1$ if areal units $(k,j)$ share a common border (are spatially close), $w_{kj}=0$ otherwise, and  $w_{kk}=0~\forall ~k$. CAR models  model  data in neighbouring  areal units ($k,j$) (those with $w_{kj}=1$) as partially autocorrelated, while those relating to non-neighbouring areal units ($k,j$) (those with $w_{kj}=0$) are assumed to be conditionally independent. Thus while the spatial autocorrelation structure implied by these CAR models depends on $\mathbf{W}$,  the appropriateness of the choice of $\mathbf{W}$ for the data at hand or the sensitivity of the results to changing its specification are rarely acknowledged or assessed in the modelling. This is in sharp contrast to the related field of geostatistics for point level data, where variogram analysis is routinely used to identify an appropriate spatial correlation structure for the data, such as assessing the validity of isotropy and choosing an appropriate parametric autocovariance model.  

Furthermore, specifying $\mathbf{W}$ based on the simple \emph{border sharing} rule is unlikely to provide an appropriate correlation structure for the count data under study,  because spatial correlation is unlikely to be present universally throughout the study region. Instead, there will be pairs of neighbouring areal units that exhibit large differences between their data values, which can be driven by complex environmental and / or social process (\citealp{mitchell2014}). Examples that illustrate this phenomenon include the fields of spatial clustering (\citealp{knorrheld2000})  and boundary analysis (\citealp{lee2012}), where identifying the locations of these step-changes is of primary interest.

Numerous approaches have been proposed for identifying spatial step-changes in areal unit count data, including specifying piecewise constant mean models (e.g. \citealp{knorrheld2000}), and modelling elements in $\mathbf{W}$ that correspond to adjacent areal units as Bernoulli random variables (\citealp{ma2010}). The latter approach allows one to estimate the spatial partial autocorrelation structure in the data, but it suffers from parameter identifiability problems because there are many more elements in $\mathbf{W}$ to estimate than there are areal units (data points). A partial solution is to propose a simple parametric model for the elements in $\mathbf{W}$ based on covariate information as in \cite{lee2012}, but the estimation of $\mathbf{W}$ is then restricted by the parametric nature of the model.

 Therefore this paper proposes a novel graph-based optimisation algorithm for estimating an appropriate neighbourhood matrix $\mathbf{W}_{E}$ for the data, which overcomes the two parameterisation issues highlighted above. The estimation of $\mathbf{W}_{E}$ is based on an initial graph  $G$, where the $K$ areal units comprise the vertex-set $V(G)$, and the edge-set $E(G)$ is defined by $\mathbf{W}$ via $E(G)=\{(k,j)|w_{kj}=1\}$ (so $\mathbf{W}$ is the adjacency matrix of $G$). The algorithm estimates whether each edge in the graph should be removed or not, with the mild restriction that every vertex must retain at least one incident edge. Our estimation algorithm has two stages, the first of which estimates $\mathbf{W}_{E}$ from the data after covariate effects have been accounted for, which is akin to using variogram analysis on detrended geostatistical data to estimate an appropriate correlation structure. The second stage of our estimation algorithm fits a Poisson log-linear model with spatio-temporally correlated random effects to the count data based on $\mathbf{W}_{E}$, with inference in a Bayesian paradigm using integrated nested Laplace Approximations (INLA, \citealp{rue2009}).  Our general Poisson log-linear count data model with CAR structured random effects  is outlined in Section 2, while our graph-based optimisation algorithm is outlined in Section 3. The superiority of our estimated $\mathbf{W}_{E}$ compared with a traditional border sharing based neighbourhood matrix $\mathbf{W}$ is thoroughly evidenced by simulation  in Section 4, while in Section 5 our approach is applied to a new respiratory disease surveillance study  based in Greater Glasgow in Scotland. Finally, Section 6 concludes the paper.

\section{Spatio-temporal areal unit modelling for count data}
The study region is partitioned into $K$  non-overlapping areal units such as Census Tracts, and data are available for each of these units for  $t=1,\ldots,N$ consecutive time periods. The outcome variable $Y_{kt}$ is a spatio-temporally aggregated count of the number of events that occur in areal unit $k$ during time period $t$, and is often accompanied by a vector of $p$ covariates $\mathbf{x}_{kt}$ and an expected count $e_{kt}$. The latter allows for the fact that the areal units have different population sizes and age-sex demographics which thus affects the observed count, and $e_{kt}$ is typically included as an offset term when modelling these data. A general model for these data within a Bayesian inferential setting  is given by

\begin{eqnarray}
Y_{kt}&\sim& \mbox{Poisson}(e_{kt}\theta_{kt})~~~~\mbox{for }k=1,\ldots,K\mbox{ and }t=1,\ldots,N,\label{likelihood}\\
\ln(\theta_{kt})&=&\mathbf{x}_{kt}^{\top}\bd{\beta} + \phi_{kt} + \delta_{t},\nonumber\\
\beta_j&\sim&\mbox{N}(0, 100000)~~~~\mbox{for }j=1,\ldots,p.\nonumber
\end{eqnarray}

Here $\theta_{kt}$ denotes the risk or rate of the outcome variable $Y_{kt}$ relative to the expected count $e_{kt}$, and the spatio-temporal variation in this risk (rate) is modelled by covariates $\{\mathbf{x}_{kt}\}$ and random effects $\{\psi_{kt} = \phi_{kt} + \delta_{t}\}$. The covariate regression parameters $\bd{\beta}=(\beta_1,\ldots, \beta_p)$ are assigned independent  weakly informative zero-mean Gaussian prior distributions with a large variance, to ensure the data play the dominant role in estimating their value. An appropriate random effects structure depends on both the residual spatio-temporal structure in the data and the goal of the analysis, and here we utilise the model proposed by \cite{waller1997} which decomposes this into separate spatial surfaces $\bd{\phi}_t=(\phi_{1t},\ldots,\phi_{Kt})$ for each time period $t$ and an overall temporal trend $\bd{\delta}=(\delta_1,\ldots,\delta_N)$. We adopt this structure because we believe that while the residual spatial surfaces will be similar each year they will not be identical. Thus assuming there is a single spatial structure common to all years as in \cite{knorrheld2000b} will be overly restrictive. We model the residual temporal trend by the first order autoregressive process:

\begin{eqnarray}
\delta_t|\delta_{t-1}&\sim&\mbox{N}\left(\alpha\delta_{t-1}, \frac{1}{\sigma}\right)~~~~\mbox{for }t=2,\ldots,N\label{ar1}\\
\delta_1&\sim&\mbox{N}\left(0, \frac{1}{\sigma(1-\alpha^2)}\right)\nonumber\\
\ln[\sigma(1-\alpha^2)]&\sim&\mbox{log-Gamma}(1, 0.00005)\nonumber\\
\ln\left(\frac{1+\alpha}{1-\alpha}\right) &\sim &\mbox{N}(0, 0.15).\nonumber
\end{eqnarray}

The prior distributions and their parameterisations are chosen to be weakly informative, and are the default specifications suggested by the INLA  software (\citealp{rue2009}) that we use for inference. We model the residual spatial trend for time period $t$ using the conditional autoregressive prior proposed by \cite{leroux2000} which is given by

\begin{eqnarray}
\phi_{kt}|\bd{\phi}_{-kt},&\sim&\mbox{N}\left(\frac{\rho_t\sum_{j=1}^{K}w_{kj}\phi_{jt}}{\rho_t\sum_{j=1}^{K}w_{kj} + 1-\rho_t}, \frac{1}{\tau_t\left[\rho_t\sum_{j=1}^{K}w_{kj}+1-\rho_t\right]}\right)\label{leroux}\\
\ln(\tau_t) &\sim & \mbox{log-Gamma}(1, 0.00005)\nonumber\\
\ln\left(\frac{\rho_t}{1-\rho_t}\right) &\sim &\mbox{N}(0, 10),\nonumber
\end{eqnarray}

where $\bd{\phi}_{-kt}=(\phi_{1t},\ldots,\phi_{k-1,t}, \phi_{k+1,t},\ldots, \phi_{kt})$. Spatial  autocorrelation is induced by the neighbourhood matrix $\mathbf{W}$, and we adopt the commonly used binary \emph{border sharing} definition described above. The level of spatial dependence is controlled globally by $\rho_t$, with $\rho_t=0$ corresponding to spatial independence (as (\ref{leroux}) simplifies to $\phi_{kt}\sim\mbox{N}(0, 1/\tau_t)$), while if $\rho_t=1$ then (\ref{leroux}) becomes the intrinsic CAR model proposed by \cite{besag1991}. A weakly-informative normal prior on the logit scale is specified for the spatial dependence parameter $\rho_t$, while a weakly informative log-gamma prior is specified for the log of the spatial precision $\tau_t$, again following the defaults suggested by the INLA software. The partial spatial autocorrelation structure implied by this model is given by

\begin{equation}
\mbox{Corr}(\phi_{kt}, \phi_{jt} |\bd{\phi}_{-kjt}) ~=~\frac{\rho_t w_{kj}}{\sqrt{\left(\rho_t\sum_{l=1}^{K}w_{kl}+1-\rho_t\right)\left(\rho_t\sum_{l=1}^{K}w_{jl}+1-\rho_t\right)}}\label{eq partial},
\end{equation}

where $\bd{\phi}_{-kjt} = \bd{\phi}_{t}\setminus\{\phi_{kt}, \phi_{jt}\}$. Thus $\mathbf{W}$ controls the partial spatial autocorrelation structure in $\bd{\phi}_t$, because if $w_{kj}=1$ then $(\phi_{kt}, \phi_{jt})$ are partially correlated with the strength of that correlation controlled globally for all pairs of neighbouring areas by $\rho_t$, whereas if $w_{kj}=0$ then $(\phi_{kt}, \phi_{jt})$ are conditionally independent. Thus while $\mathbf{W}$ is  crucial to the model because it determines the spatial correlation structure in the data, its appropriateness for the data or the sensitivity of the results to changing its specification are rarely assessed. Furthermore, specifying $\mathbf{W}$ via border sharing implies that all pairs of geographically adjacent areal units will have correlated random effects, which spatially smooths their values towards each other. However, the residual spatial surface in real data sets often exhibit areas of spatial smoothness separated by step changes, an example of which  can be seen in Figure \ref{figSMR}.  Additionally, the identification of such step changes can be the goal of the analysis, such as in the areas of spatial clustering and boundary analysis highlighted earlier. Therefore in the next section we propose a novel graph-based optimisation algorithm for estimating a more appropriate neighbourhood matrix $\mathbf{W}_E$ for the data that leads to improved inference.

\section{Methodology}
We propose a novel two-stage approach for estimating the model parameters\\ 
 $\bd{\Theta}=(\bd{\beta},\bd{\delta}, \sigma, \alpha, \bd{\phi}_1,\ldots,\bd{\phi}_N,  \rho_1,\ldots,\rho_N, \tau_1,\ldots,\tau_N)$ and an appropriate neighbourhood matrix $\mathbf{W}_{E}$, which extends the currently used approach of naively fixing $\mathbf{W}$ based on the border sharing rule.  In stage 1 we estimate $\mathbf{W}_{E}$ using a graph-based optimisation  algorithm, and in stage 2 we estimate the  posterior distribution $f(\bd{\Theta} |\mathbf{W}_{E}, \mathbf{Y})$ conditional on $(\mathbf{Y}, \mathbf{W}_{E})$. Our methodology thus brings areal unit modelling into line with standard practice in geostatistical modelling, which is to first estimate a trend model and then identify an appropriate correlation structure via residual analysis. In what follows $\mathbf{W}$ denotes the neighbourhood matrix constructed based on border sharing, while $\mathbf{W}_{E}$ denotes our estimated matrix.

\subsection{Stage 1 - Estimating $\mathbf{W}_{E}$}
We estimate a single $\mathbf{W}_{E}$ for the data, which requires the residual spatial structure to be similar for all time periods. We do this because we need multiple realisations of the spatial surface to estimate its correlation structure via $\mathbf{W}_E$ well, which is  evidenced by the simulation study in Section 4. Therefore, first we estimate a single residual spatial surface $\tilde{\bd{\phi}}=(\tilde{\phi}_1,\ldots,\phi_K)$ for all time periods that is used to estimate $\mathbf{W}_{E}$.

\subsubsection{Estimating $\tilde{\bd{\phi}}$}
In classical geostatistics with normally distributed data and mean model $\ex{Y_{kt}}=\mathbf{x}_{kt}^{\top}\bd{\beta}$, one examines the raw residuals $\tilde{\phi}_{kt}=Y_{kt}-\mathbf{x}_{kt}^{\top}\hat{\bd{\beta}}$ to identify an appropriate correlation structure, where initially $\hat{\bd{\beta}}$ is estimated assuming independent errors. The analogous approach for our count data model (\ref{likelihood})  rearranges $\ex{Y_{kt}}=e_{kt}\exp(\mathbf{x}_{kt}^{\top}\bd{\beta} + \phi_{kt} + \delta_t)$  to give

\begin{equation}
\tilde{\phi}_{kt}=\ln\left(\frac{\ex{Y_{kt}}}{e_{kt}}\right)-\mathbf{x}_{kt}^{\top}\bd{\beta} - \delta_t ~\approx~\ln\left(\frac{Y_{kt}}{e_{kt}}\right)-\mathbf{x}_{kt}^{\top}\hat{\bd{\beta}}.\label{resids}
\end{equation}

This replaces the unknown $\ex{Y_{kt}}$ with the observed data $Y_{kt}$. The mean model parameters $\bd{\beta}$ are again estimated assuming independence, and $\{\delta_t\}$ is removed as it is constant over space and hence does not impact on the estimation of the spatial correlation structure. Then we estimate a common residual spatial surface by averaging over the $N$ time periods, that is $\tilde{\phi}_k=(1/N)\sum_{t=1}^{N}\tilde{\phi}_{kt}$ for all $k$.

\subsubsection{Deriving an objective function to optimise}
The CAR model (\ref{leroux}) represents a graph $G$ whose vertex-set $V(G)$ is the set of $K$ areal units, and whose edge-set is $E(G)=\{(k,j)|w_{kj}=1\}$, a subset of un-ordered pairs of elements of $V(G)$. In graph theoretic terms $G$ is the simple graph with adjacency matrix $\mathbf{W}$, where $\mathbf{W}=(w_{kj})$ is defined by the border sharing rule.  Given $\tilde{\bd{\phi}}$ we estimate $\mathbf{W}_{E}$ by searching for a suitable subgraph of $G$ which maximises the value of an objective function $J(\tilde{\bd{\phi}})$. We base the objective function on the special case of (\ref{leroux}) where $\rho_t=1$, which assumes that all pairs of neighbouring areal units (those with $w_{kj}=1$) have correlated random effects. This allows step changes in the spatial surface to be identified  by removing edges from the graph (e.g. setting $w_{E_{kj}}=0$), which would make the corresponding random effects conditionally independent as illustrated by (\ref{eq partial}). Therefore fixing $\rho_t=1$ in (\ref{leroux}) and dropping the time subscript $t$ as $\tilde{\bd{\phi}}$ is an average over all time periods,  we obtain the following objective function.

\begin{equation}
J(\tilde{\bd{\phi}}) ~=~\ln\left[\prod_{k=1}^{K}f(\tilde{\phi}_{k}|\tilde{\bd{\phi}}_{-k})\right] ~=~\ln\left[\prod_{k=1}^{K}\mbox{N}\left(\frac{\sum_{j=1}^{K}w_{kj}\tilde{\phi}_{j}}{\sum_{j=1}^{K}w_{kj}},~
\frac{1}{\tau\left[\sum_{j=1}^{K}w_{kj}\right]}\right)\right].
\end{equation}

After removing unnecessary constants  $J(\tilde{\bd{\phi}}) $ becomes

\footnotesize
\begin{eqnarray}
J(\tilde{\bd{\phi}})&\propto&\frac{K}{2}\ln\left(\tau\right) + \frac{1}{2}\sum_{k=1}^{K}\ln\left(\sum_{j=1}^{K}w_{kj}\right) - \frac{\tau}{2}\sum_{k=1}^{K}\left(\sum_{j=1}^{K}w_{kj}\right)\left(\tilde{\phi}_k - \frac{\sum_{r=1}^{K}w_{kr}\tilde{\phi}_{r}}{\sum_{r=1}^{K}w_{kr}}\right)^{2},\label{eqnobjective1}
\end{eqnarray}
\normalsize

which depends on the precision parameter $\tau$. Estimating $\tau$ by maximising (\ref{eqnobjective1}) yields the maximum likelihood estimator $\hat{\tau}= K / \sum_{k=1}^{K}\left(\sum_{j=1}^{K}w_{kj}\right)\left(\tilde{\phi}_k - \frac{\sum_{r=1}^{K}w_{kr}\tilde{\phi}_{r}}{\sum_{r=1}^{K}w_{kr}}\right)^{2},$ which when plugged into (\ref{eqnobjective1}) yields the final objective function

\footnotesize
\begin{eqnarray}
J(\tilde{\bd{\phi}})&\propto&  \frac{1}{2}\sum_{k=1}^{K}\ln\left(\sum_{j=1}^{K}w_{kj}\right)-\frac{K}{2}\ln\left[\sum_{k=1}^{K}\left(\sum_{j=1}^{K}w_{kj}\right)\left(\tilde{\phi}_k - \frac{\sum_{r=1}^{K}w_{kr}\tilde{\phi}_{r}}{\sum_{r=1}^{K}w_{kr}}\right)^{2}\right].\label{eqnobjective}
\end{eqnarray}
\normalsize

This function only depends on $(\tilde{\bd{\phi}}, \mathbf{W})$, where the latter is the only thing to be maximised as $\tilde{\bd{\phi}}$ is estimated as described above.

\subsubsection{Graph-based optimisation}
Let $H$ be generic notation for any graph, then we use the following graph theoretic terminology in this section: (i) we write $uv$ for the edge $\{u,v\}$ with endpoints $u$ and $v$; (ii) an edge $e \in E(H)$ is said to be \emph{incident with} a vertex $v \in V(H)$ if $v$ is an endpoint of $e$; (iii) the number of edges in $H$ incident with any single vertex $v$, written $\deg_H(v)$, is called the \emph{degree} of $v$ in $H$; (iv) we write $N_H(v)$ for the set $\{u \in V(H) \setminus\{v\}: uvv \in E(H)\}$ of \emph{neighbours} of $v$ in $H$; (v) a graph $H'$ is a \emph{subgraph} of $H$ if $V(H') \subseteq V(H)$ and $E(H') \subseteq E(H)$; and  (vi) if $H$ and $H'$ have the same vertex set we say that $H'$ is a \emph{spanning subgraph} of $H$. 

The graph $G$  based on $\mathbf{W}$ has vertex-set $V(G)$ and edge-set $E(G)$, and we assume that edges $e \in E(G)$ can be removed from the graph but that new edges cannot be added in. This means that one can estimate $w_{E_{kj}}=\{0,1\}$ if $w_{kj}=1$, but if $w_{kj}=0$ then $w_{E_{kj}}$ remains fixed at zero. Additionally, we assume that each area (vertex) must retain at least one  edge in the graph, which corresponds to the constraint $\sum_{j=1}^{K}w_{E_{kj}}>0$ for all $k$. This ensures that we do not divide by 0 in (\ref{eqnobjective}). Let $f(H,\tilde{\bd{\phi}})$ denote the value of $J(\tilde{\bd{\phi}})$ corresponding to $\mathbf{W}_H$, the adjacency matrix corresponding to the sub-graph $H$ of $G$. Then the goal of our optimisation problem can be phrased as finding a spanning subgraph $\tilde{G}$ of $G$, with minimum degree at least one, which maximises $f(\tilde{G}, \tilde{\bd{\phi}})$.  

This graph optimisation problem is known to be NP-hard (\cite{lee2020complexity}), and so is extremely unlikely to admit an exact algorithm which will terminate in polynomial time on all possible inputs.  Moreover, this intractability result holds even if we assume that the input graph $G$ is planar; our input graph is necessarily planar because it is derived from the adjacencies of non-overlapping regions in the plane.  In this work we therefore adopt a heuristic local search approach, which we describe in detail in the rest of this section.  It should be emphasised that this algorithm is not guaranteed to find the global optimal solution; we leave a more in-depth study of the existence or otherwise of algorithms with provable performance guarantees for future work.

A brute force optimisation strategy would consider all possible subsets of edges to delete (which is exponential in the number of edges in the original graph), and choose the one which maximises the objective function. However such a running-time is already infeasible in our relatively small example with $671$ edges.  To avoid this, we instead obtain an improved matrix $\mathbf{W}_{E}$ by carrying out a sequence of local optimisation operations; this is much faster, but is not guaranteed to result in a globally optimal solution.

For our heuristic local optimisation, we consider the vertices of the graph in some fixed order, and attempt to optimise the set of edges incident with each vertex in turn.  The reason that this does not necessarily find a global optimum is that the effect of deleting the edge $uv$ depends on the set of edges incident at both $u$ and $v$, so we have to choose a set of neighbours to retain for $v$ without necessarily knowing which neighbours $u$ will retain in the final solution.  To deal with this, we decide whether or not to delete an edge by considering the difference between the contribution to the objective function from $u$ (respectively $v$) from the best possible set of incident edges at $u$ (respectively $v$) that does include the edge $uv$, and the best possible set that does not include this edge.

In order to apply this strategy, we need to express the objective function as a sum of contributions associated with each vertex of the graph, so that we can assess the impact of making local changes associated with an individual vertex.  As a first step, we reformulate equation \eqref{eqnobjective} in more graph theoretic notation.  To do this, we set $V = V(G)$ (observing that we use the same vertex set throughout), and note that $|V| = K$.  For a vertex $v$ corresponding to region $k$ in the matrix, we set $\tilde{\phi}_v = \tilde{\phi}_k$.  This gives
\begin{equation}\label{eqn:obj-fn-graph}
f(H,\tilde{\bd{\phi}}) \propto  \frac{1}{2}\sum_{v \in V} \ln\left(\deg_H(v)\right)-\frac{K}{2}\ln\left[\sum_{v \in V} \deg_H(v) \left(\tilde{\phi}_v - \frac{\sum_{u \in N_H(v)} \tilde{\phi}_u}{\deg(v)}\right)^{2}\right].
\end{equation}
To simplify notation, we will write $\ND_H(v,\tilde{\bf{\phi}})$ for the \emph{neighbourhood discrepancy} defined as $\left(\tilde{\phi}_v - \frac{\sum_{u \in N_H(v,\tilde{\bf{\phi}})} \tilde{\phi}_u}{\deg_H(v)}\right)^{2}$.  It is now clear that, to maximise the right-hand side of \eqref{eqn:obj-fn-graph}, on the one-hand we would like to retain as many edges as possible to maximise the first term, but on the other hand we minimise the second term by deleting edges to decrease the neighbourhood discrepancy at each vertex. We can now associate with a given vertex $v$ the following contribution, $\cont(v,H,\tilde{\bd{\phi}})$, to the right-hand side of equation \eqref{eqn:obj-fn-graph}:
\begin{align*}
\cont(v,H,\tilde{\bd{\phi}}) & := \frac{\ln (\deg_H(v))}{2} - \frac{K}{2} \ln \left[ \sum_{w \in V} \deg_H(w)\ND_H(w,\tilde{\bf{\phi}})\right] \\
	& \qquad \qquad + \frac{K}{2}\ln\left[ \sum_{w \in V \setminus \{v\}} \deg_H(w) \ND_H(w,\tilde{\bf{\phi}})\right]\\
	& = \frac{\ln (\deg_H(v))}{2} - \frac{K}{2} \ln \left[\sum_{w \in V \setminus \{v\}} \deg_H(w)\ND_H(w,\tilde{\bf{\phi}}) + \deg_H(v)\ND_H(v,\tilde{\bf{\phi}})\right] \\
	& \qquad \qquad + \frac{K}{2} \ln \left[\sum_{w \in V \setminus \{v\}} \deg_H(w)\ND_H(w,\tilde{\bf{\phi}})\right] \\
	& = \frac{\ln (\deg_H(v))}{2} - \frac{n}{2} \ln \left[1 + \frac{\deg_H(v)\ND_H(v,\tilde{\bf{\phi}})}{\sum_{w \in V \setminus \{v\}} \deg_H(w)\ND_H(w,\tilde{\bf{\phi}})}\right].
\end{align*}
We then have that $f(H,\tilde{\bd{\phi}}) \propto \sum_{v \in V} \cont(v,H,\tilde{\bd{\phi}})$.

The remaining barrier to using this expression to carry out locally optimal modifications is that the value of $\sum_{w \in V \setminus \{v\}} \deg_H(w) \ND_H(w,\tilde{\bf{\phi}})$ depends on the entire graph, not just the edges incident with $v$, so we cannot compute the value of $\cont(v,H,\tilde{\bf{\phi}})$ knowing only the neighbours of $v$ in $H$.  To deal with this, we define the \emph{adjusted contribution} of $v$ in $H$, with respect to a second graph $H'$:
\begin{align*}
\adjcont_{H'}&(v,H,\tilde{\bd{\phi}}) := \frac{\ln (\deg_H(v))}{2}\\
& - \frac{n}{2} \ln \left[1 + \frac{\deg_H(v)\ND_H(v,\tilde{\bf{\phi}})}{\sum_{w \in V} \deg_{H'}(w)\ND_{H'}(w,\tilde{\bf{\phi}}) - \deg_H(v)\ND_H(v,\tilde{\bf{\phi}})}\right].
\end{align*}
Observe that, if $H$ is a spanning subgraph of $H'$, we have $\sum_{v \in V} \ln(\deg_H(v)) \le \sum_{v \in V} \ln(\deg_{H'}(v))$ and so, if $f(H,\tilde{\bf{\phi}}) > f(H',\tilde{\bf{\phi}})$, we must have
\begin{align*}
\sum_{w \in V \setminus \{v\}} & \deg_H(w)\ND_H(w,\tilde{\bf{\phi}}) < \sum_{w \in V} \deg_{H'}(w)\ND_{H'}(w,\tilde{\bf{\phi}}) - \deg_H(v)\ND_H(v,\tilde{\bf{\phi}}).
\end{align*}
This tells us that, if $\adjcont_H(v,H \setminus \{e\},\tilde{\bf{\phi}}) > \adjcont_H(v,H,\tilde{\bf{\phi}})$, then the contribution at $v$ is still increased by deleting $e$ even when deletions are also carried out elsewhere in the graph to decrease the weighted sum of neighbourhood discrepancies.

These observations motivate our iterative approach.  At the first step we consider the first vertex $v$ and use the original graph $G$ to identify a set of edges incident with $v$ to delete (by considering the adjusted contribution with respect to $G$ at both endpoints of the edges in question).  We then delete these edges to obtain a new graph $G'$ and continue with the next vertex, this time considering the adjusted contribution with respect to $G'$.  We continue in this way, returning to the start of the vertex list when we reach the end, until we complete a pass through all remaining feasible vertices (that is, those which still have more than one neighbour in the modified graph) without identifying any deletions that increase the objective function.

The algorithm is summarised in pseudocode as Algorithm \ref{alg:local-search} in the appendix.  We note that the running-time depends exponentially on the maximum degree, but only linearly on the number of edges. It is not unreasonable to expect that the maximum degree will in practice be small compared with the total number of vertices or edges: it is unlikely that any one areal unit will border a very large number of other units (in our example the maximum degree is $22$).  Software in the form of a suite of \texttt{Python} functions (which can be incorporated into \texttt{R} using the \texttt{reticulate} package) to implement the optimisation are available at\\ \texttt{https://github.com/kittymeeks/spatial-stats-optimisation}.

\subsection{Stage 2 - Estimating $\bd{\Theta}$ given $\mathbf{W}_{E}$}
We fit model (\ref{likelihood}) - (\ref{leroux}) with $\mathbf{W}_{E}$ replacing $\mathbf{W}$ in a Bayesian setting  using integrated nested Laplace approximations (INLA, \citealp{rue2009}) using the full Laplace approximation. We use INLA due to its computational speed in fitting the models, but we could have used Markov chain Monte Carlo (MCMC) simulation methods, for example using the \texttt{CARBayesST} package in \texttt{R} written by \cite{lee2018}.

\section{Simulation study}
 This section presents a simulation study that compares the  performance of model (\ref{likelihood}) - (\ref{leroux}) based on a neighbourhood matrix that is: (i) constructed using the border sharing rule (denoted by $\mathbf{W}$); or (ii) estimated using graph-based optimisation (denoted by $\mathbf{W}_E$).

 \subsection{Data generation}
The study region is the $K=257$ Intermediate Zones (IZ) that make up the Greater Glasgow and Clyde Health Board (GGCHB) in Scotland, which is the setting for the motivating case study presented in Section 5. Count data are generated for this region from model (\ref{likelihood}), and we consider scenarios with  $N=1,5,9$ time periods to see how this affects the performance of our methodology. We also examine how the size of the counts $\{Y_{kt}\}$ affects estimation performance, by considering scenarios where the  expected counts $\{e_{kt}\}$ are drawn uniformly within the ranges: (i) $[10, 30]$ (rare events); and (ii)  $[150, 250]$ (common events). Finally, we also vary the sizes of the step changes we generate in the residual surface $\bd{\phi}_t$. 

Each simulated data set includes an independent ($\mathbf{x}_1$) and a spatially autocorrelated ($\mathbf{x}_2$) covariate, and the corresponding regression parameters are fixed at $\beta_1=\beta_2=0.05$. Both covariates are generated from zero-mean multivariate normal distributions with a standard deviation of 0.5 separately for each time period, with the independent covariate $\mathbf{x}_1$ having the identity correlation matrix. The correlation matrix for $\mathbf{x}_2$ is defined by the spatial exponential correlation matrix $\bd{\Sigma}=\exp(-\xi \mathbf{D})$,  where $\mathbf{D}$ is a $K\times K$ distance matrix between the centroids of the $K$ IZs. The spatial range parameter $\xi$ was chosen to ensure the covariate was visually spatially smooth,  which was achieved by fixing $\xi$ so that the mean correlation across all pairs of IZs was 0.25.

Temporal autocorrelation was induced into each simulated data set by a first order autoregressive process, with AR(1) coefficient $\alpha=0.8$. Similarly, spatial autocorrelation was induced via a multivariate normal distribution with a  spatial exponential correlation matrix $\bd{\Sigma}=\exp(-\xi \mathbf{D})$, where $\xi$ was chosen so that the mean pairwise correlation across all IZs was 0.15. To ensure that each time period had a similar but not identical residual spatial surface, $\bd{\phi}_t$ was generated  by the sum $\bd{\phi}_t=\bd{\phi} + \bd{\phi}^{*}_t$, with a common spatial surface $\bd{\phi}$ for all time periods and time period specific deviations $\bd{\phi}^{*}_t$ with a lower variance. The mean of $\bd{\phi}_t$ is denoted by  $\bd{\mu}$, and this is the mechanism by which step changes are induced into $\bd{\phi}_t$. Specifically, $\bd{\mu}$ is piecewise constant with levels $(-\lambda, 0, \lambda)$, where $\lambda$ determines the size of the step changes. Here we consider values of $\lambda=0,0.25,0.5$ in our simulation design, where $\lambda=0$ corresponds to no step changes while $\lambda=0.5$ corresponds to large step changes. These mean values $(-\lambda, 0, \lambda)$ are assigned to the IZs to match the  structure of the case study data as closely as possible, with for example IZs that exhibit comparatively high  rates $\{\theta_{kt}\}$ being assigned a mean value of $\lambda$. Example realisations of $\bd{\phi}_t$ for all 3 values of $\lambda$ are presented in Section 1 of the supplementary material accompanying this paper.

\subsection{Results}
One hundred data sets are generated under  each of 18 scenarios, which include all possible combinations of: (i) $N=1,5,9$; (ii)  $e_{kt}\in[10, 30], [150, 250]$; and (iii) and $\lambda=0,0.25,0.5$. The accuracy of the risk (rate) estimates $\{\hat{\theta}_{kt}\}$ are summarised here, because they are the ones of greatest interest in the motivating study presented in the next section. In contrast, the accuracy of  the covariate effect estimates $\hat{\bd{\beta}}$ are presented in Section 2 of the supplementary material. The accuracy of each approaches estimates $\{\hat{\theta}_{kt}\}$ are summarised in Table \ref{sim_theta}, which displays their root mean square errors (RMSE) as well as the coverage probabilities and average widths of the associated 95\%  credible intervals. 

The table shows 3 main findings, the first of which is that if you have purely spatial data  ($N=1$), then estimating $\mathbf{W}_{E}$ leads to worse results than using the simple border sharing matrix $\mathbf{W}$. This worse performance is highlighted by slightly larger RMSEs  and reduced coverage probabilities below the nominal 95\% levels. This worse performance occurs because the estimate of $\tilde{\bd{\phi}}$ used in the objective function (\ref{eqnobjective}) is only based on one set of spatial residuals from (\ref{resids}), and thus does not provide a good enough estimate of the unknown residual structure in the data. Secondly, if $N>1$ but the risk surface does not exhibit step changes ($\lambda=0$), then the RMSEs are broadly similarly between the two methods. However, the uncertainty quantification is better when using $\mathbf{W}_{E}$, with coverage probabilities closer to 95\% when the disease is rare ($e_{kt}\in[10, 30]$) and narrower intervals with similar coverage probabilities when the disease is common ($e_{kt}\in[150, 250]$).

Finally, if one has spatio-temporal data ($N>1$) that contain step changes ($\lambda>0$), then using $\mathbf{W}_{E}$ always produces better risk (rate) estimation compared with using $\mathbf{W}$. This improved estimation includes reduced RMSEs by between 11.8\% and 25.9\%, and similar coverage probabilities obtained from credible intervals that are narrower by between 10.3\% and 22.1\%. Both these improvements occur because $\mathbf{W}_{E}$ better represents the residual spatial structure in the data than $\mathbf{W}$, such as allowing for the locations of step changes by setting the appropriate $w_{E_{kj}}=0$. As the data contain multiple time periods the replication in the spatial surface leads to better estimates of  $\tilde{\bd{\phi}}$ compared to when $N=1$, which causes the improvements in inference. The reduced widths of the 95\% credible intervals when using $\mathbf{W}_{E}$ is because this matrix does not enforce correlation between neighbouring areas that exhibit a step change between them. This means that the variance $1/\tau$ is not inflated to account for the spatial smoothing that is enforced between those areal units with very different data values.

\begin{table}
\caption{Accuracy of the estimated risks (rates) $\{\theta_{kt}\}$ from the model with the border sharing ($\mathbf{W}$) and estimated ($\mathbf{W}_{E}$) neighbourhood matrices.\label{sim_theta}}
\begin{tabular}{llrrrr}\hline
\textbf{Step}&\textbf{Time}&\multicolumn{4}{c}{\textbf{Disease prevalence}}\\
\multicolumn{1}{l}{\textbf{change}}&\multicolumn{1}{l}{\textbf{periods}}& \multicolumn{2}{c}{$\mathbf{e_i\in[10, 30]}$}&\multicolumn{2}{c}{$\mathbf{e_i\in[150, 250]}$}\\\hline
\textbf{RMSE}&&$\mathbf{W}$ &$\mathbf{W}_E$ & $\mathbf{W}$ &$\mathbf{W}_E$\\
&$N=1$ &0.084 &0.110 &0.048 & 0.054\\
$\lambda=0$& $N=5$& 0.091 &0.091 &0.052 &0.049\\
& $N=9$ &0.092 &0.090 &0.052 &0.048 \\\hline
&$N=1$ &0.150 &0.168 &0.067 &0.066\\
$\lambda=0.25$&$N=5$ &0.152 &0.134 &0.068 &0.057\\
&$N=9$ &0.153 &0.129 &0.069 &0.056\\\hline
&$N=1$ &0.204 &0.208 &0.073 &0.071 \\
$\lambda=0.5$&$N=5$ &0.205 &0.163 &0.075 &0.061\\
&$N=9$ &0.205 &0.152 &0.079 &0.060\\\hline
\textbf{Coverage}&&$\mathbf{W}$ &$\mathbf{W}_{E}$ & $\mathbf{W}$ &$\mathbf{W}_{E}$\\
\textbf{(width)}&$N=1$ &62.8 (0.147) &95.5 (0.428) &94.6 (0.187) &88.7 (0.174)\\
$\lambda=0$&$N=5$ &82.3 (0.287) &93.5 (0.347)  &94.6 (0.201) & 94.1 (0.186)\\
&$N=9$ &83.1 (0.292) &88.5 (0.320) &94.6 (0.204) &94.6 (0.187)\\\hline
&$N=1$ &94.7 (0.579) &89.1 (0.541) &95.1 (0.260) &90.7 (0.223)\\
$\lambda=0.25$&$N=5$ &94.7 (0.593) &94.8 (0.532) &94.8 (0.262) &95.3 (0.226)\\
&$N=9$ &94.7 (0.590) &95.6 (0.523) &94.9 (0.269) & 95.7 (0.227)\\\hline
&$N=1$ &94.6 (0.769) &89.1 (0.655) &94.9 (0.279) &92.2 (0.247)\\
$\lambda=0.5$&$N=5$ &94.8 (0.771) & 95.4 (0.636) &95.2 (0.289) &95.4 (0.239)\\
&$N=9$ &94.7 (0.775) &96.1 (0.618)  &95.4 (0.307) &95.7 (0.239)\\\hline
\end{tabular}
\end{table}

\section{Motivating study - respiratory ill health in Glasgow}
Health care in Scotland is managed locally by 14 regional health boards, and here we focus on the Greater Glasgow and Clyde health board (GGCHB) because it exhibits some of the poorest health and widest health inequalities in western Europe (\citealp{walsh2016}). Specifically, the health board are interested in: (i) identifying areas that exhibit elevated risks of ill health allowing the appropriate targeting of health interventions; and (ii)  quantifying whether inequalities in risk between rich and poor communities are widening or narrowing over time. We address these questions in the context of respiratory disease because it is one of the leading causes of death in Scotland (\url{https://www.nrscotland.gov.uk/statistics-and-data}).

\subsection{Data available}
Data are available on the yearly numbers of respiratory hospitalisations (ICD-10 codes J00 - J99) between 2011 and 2017 for each of the $K=257$ Intermediate Zones (IZ) that make up the GGCHB, which is a Scottish government developed small-area geography with an average population of around 4,000 people. These yearly disease counts $\{Y_{kt}\}$ are accompanied by expected counts $\{e_{kt}\}$ computed using indirect standardisation, which allow for the varying population demographics between IZs. The commonly used exploratory estimate of disease risk $\theta_{kt}$ is the standardised morbidity ratio (SMR) computed as SMR$_{kt}=Y_{kt} / e_{kt}$, and SMRs that are respectively greater / less than one indicate IZs  that exhibit  respectively higher / lower risk than the Scottish average over the study duration.

\begin{figure}
	\centering
	\begin{picture}(10,16)
	\put(-1.5,10){\scalebox{0.25}{\includegraphics{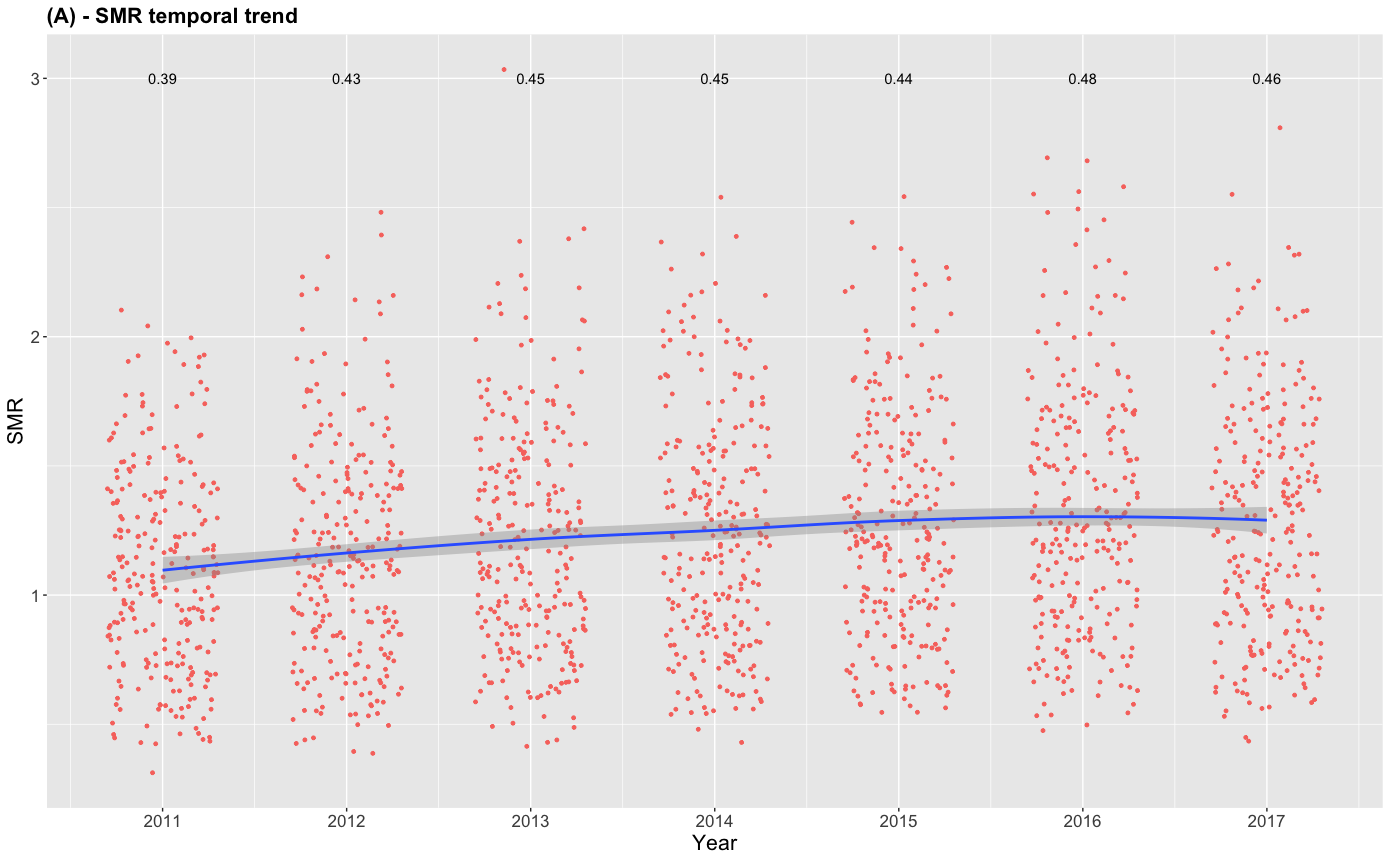}}}
	\put(-1,0){\scalebox{0.25}{\includegraphics{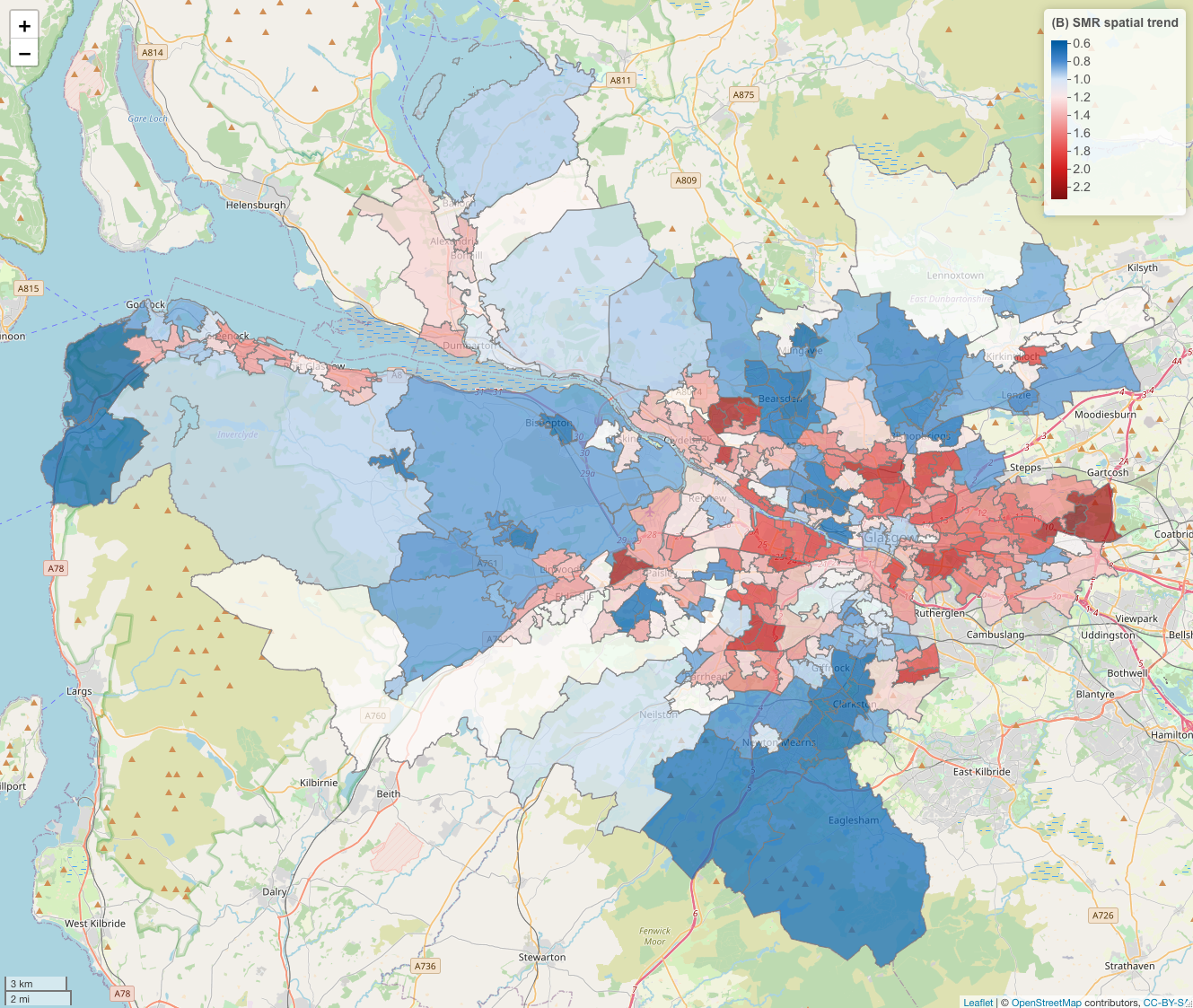}}}
	\end{picture}
	\caption{Summary of the temporal (A, top) and spatial (B, bottom) pattern in the SMR. In panel (A) the  SMR values have been jittered in the horizontal (Year) direction to improve the presentation, and the blue line is a LOESS trend. In panel (B) the total SMR over all 7 years is presented.  \label{figSMR}}
\end{figure}

The temporal (A) and spatial (B) trends in the SMR are displayed in Figure \ref{figSMR}, where in panel (A) jittering has been added to the Year direction to improve the visibility of the points, and a trend line has been estimated using LOESS smoothing. Additionally, the numbers in the plot are spatial standard deviations in the SMR, which give an idea of the changing level of health inequality over time. The figure shows a small increasing trend in the SMR over time, with average SMRs of 1.10 (a 10\% increased risk) in 2011 and 1.28 in 2017. There also appears to be a slight increase in the health inequality across the GGCHB over the 7 years, as the spatial standard deviation of the SMR increases from 0.39 in 2011 to 0.46 in 2017. The bottom panel displays the spatial pattern in the overall SMR across the 7-year period (i.e. $SMR_k=\sum_{t=1}^{7}Y_{kt} / \sum_{t=1}^{7}e_{kt}$), which shows substantial variation with SMRs ranging between 0.57 and 2.33.

We have access to a number of covariates to explain this spatial pattern in disease risk, the most important of which is the Scottish Index of Multiple Deprivation (SIMD, \url{http://www.gov.scot/Topics/Statistics/SIMD}). Deprivation or poverty is a key driving factor in spatial studies of population level ill health (\citealp{NHS2016}), in part because of its links to smoking. The SIMD is not computed each year using the same methodology,  so here we use the index  for 2016 as a purely spatial covariate. The SIMD is a composite index comprising indicators relating to access to services, crime, education, employment, health, housing, and income, and we consider each of these as possible covariate except for health as our outcome variable is health related. Furthermore, the crime indicator has one very large outlier (it is a city centre IZ containing lots of bars), so it is replaced by the average value from its neighbouring IZs. Finally, the income, employment, and education domains are all collinear, having pairwise correlations between 0.87 and 0.98. 

We also consider a measure of fine particulate matter air pollution called PM$_{2.5}$, because existing studies have shown that it is associated with respiratory ill health in Scotland (\citealp{lee2019}). In common with the above study we utilise modelled concentrations from the Pollution Climate Mapping (PCM) model (\url{https://uk-air.defra.gov.uk/data/pcm-data}), because measured data are not available at the small area IZ scale. The model produces annual average concentrations on a 1$km^2$ grid across the United Kingdom, which we spatially realign to our IZ scale by averaging.

\subsection{Stage 1 - Estimating  $\mathbf{W}_{E}$}
We first fit a simple mean model (model (\ref{likelihood}) with no random effects) to estimate the residual spatial structure in the data via (\ref{resids}). Initially, we included the three collinear SIMD indicators (education, employment and income) in separate models, and the model with education had the lowest AIC and was thus retained. The remaining covariates crime, housing, access and PM$_{2.5}$ were then added to the model, and with the exception of housing they all exhibited significant effects at the 5\% level and were used in the final mean model. This covariate only model exhibits substantial overdisperion with respect to the Poisson assumption ($\mbox{Var}(Y_{kt})=3.20\times  \ex{Y_{kt}}$), and the residuals from (\ref{resids}) exhibit substantial spatial autocorrelation, with p-values against the null hypothesis of independence based on a Moran's I permutation test being less than 0.05 for 6 out of the 7 years. 

\begin{figure}
	\centering
\scalebox{0.18}{\includegraphics{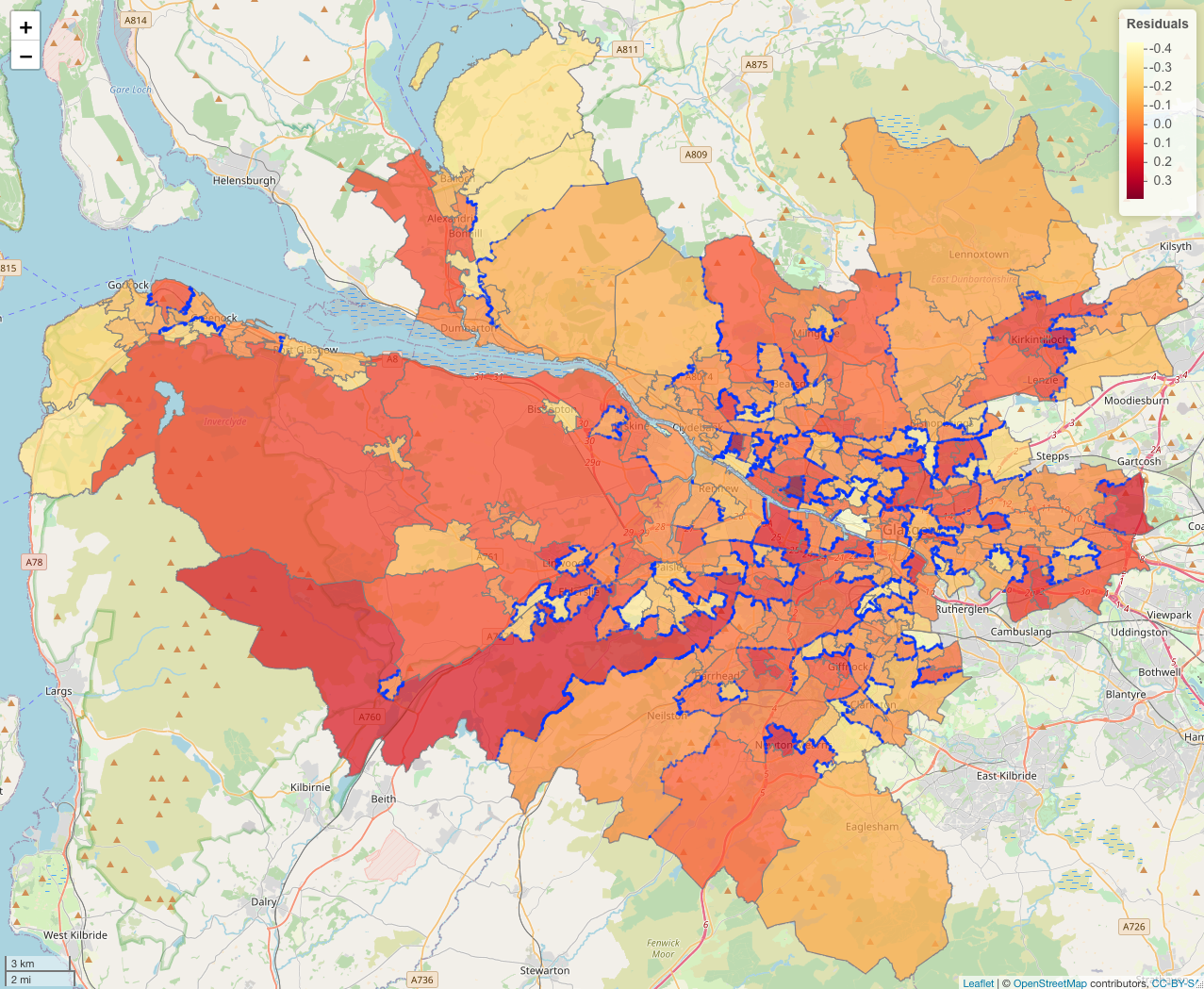}}
	\caption{Spatial map of the temporally averaged residuals from the covariate only model. The blue dots represent edges that have been removed from the graph in estimating $\mathbf{W}_E$. \label{figresid}}
\end{figure}

We then estimated $\mathbf{W}_{E}$ as described in Section 3, based on the temporally averaged residuals and $\mathbf{W}$ constructed using the border sharing rule. This initial $\mathbf{W}$ splits the $K$ IZs into two disconnected sub-graphs north and south of the  river Clyde, while the estimated graph structure $\mathbf{W}_{E}$ consists of one main sub-graph each side of the river with 4 further smaller disconnected sub-graphs. The  graph based on $\mathbf{W}$ contains 671 edges compared to 332 for $\mathbf{W}_{E}$, a 50.5\% reduction in the number of edges. The locations of the edges that have been removed are displayed as blue dots in Figure \ref{figresid}, which also displays the temporally averaged residuals across all 7 years. The figure shows that visually the residuals do not exhibit a spatially smooth surface, and that the removed edges (blue dots)  mainly correspond to locations  where there appear to be step changes. Although, note that removing an edge makes the corresponding data values conditionally and not marginally independent.

\subsection{Stage 2 - Modelling the data}
Model (\ref{likelihood}) - (\ref{leroux}) is then fitted to the data separately using $\mathbf{W}$ and $\mathbf{W}_{E}$, and inference is based on integrated nested Laplace approximations using the full Laplace approximation. A summary of the overall fit of each model via the deviance information criterion (DIC) and the effective number of independent parameters (p.d) is presented in Table \ref{modsummary}, together with other key model parameters. The table shows that the model using $\mathbf{W}_E$ fits the data better than that using $\mathbf{W}$, with reductions in the DIC of around 213 and in the p.d by around 160. The latter suggests that $\mathbf{W}_E$ provides a more parsimonious description of the data, which is due to an increase in the precisions $(\tau_1,\ldots,\tau_N)$ (summarised by the range of the posterior medians in Table \ref{modsummary})  when using $\mathbf{W}_E$. These increased precisions occur because unlike $\mathbf{W}$,  $\mathbf{W}_{E}$ does not include edges between pairs of geographically adjacent IZs that exhibit large differences in their residuals, which reduces the amount of variation between $\phi_{kt}$ and its spatially weighted mean from (\ref{leroux}). This also increases the amount of spatial dependence in each spatial surface, which can be seen by the large increases in $(\rho_1,\ldots,\rho_N)$ when using $\mathbf{W}_{E}$. 

\begin{table}\centering
\caption{Summary of the models using $\mathbf{W}$ and $\mathbf{W}_{E}$, including overall model fit (DIC) and other key model parameters. For $(\tau_t, \rho_t)$ the table displays the range in the posterior medians over time.\label{modsummary}}
\begin{tabular}{lrr}\hline
    \textbf{Quantity} & $\mathbf{W}_{BS}$ &   $\mathbf{W}_{E}$ \\\hline
DIC &14,309 &14,096\\
p.d &1,139 & 979\\\hline
Precision $\tau_t$ &28.10 - 53.39 & 38.10 - 56.80\\
Dependence $\rho_t$ &0.006 - 0.188 &0.683 - 0.763\\\hline
Access &0.989 (0.979, 1.000) & 0.983 (0.973, 0.993)\\
Crime & 0.981 (0.970, 0.992) & 0.985 (0.974, 0.996)\\
Education &1.373 (1.359, 1.387) & 1.379 (1.365, 1.394) \\
PM$_{2.5}$ & 1.020 (1.003, 1.037) &1.004 (0.986, 1.023)\\\hline
\end{tabular}
\end{table}

\subsubsection{Covariate effects}
Estimated relative risks (posterior medians) and 95\% credible intervals for the covariates are also displayed in Table \ref{modsummary}, where each relative risk relates to a one standard deviation increase in the covariates value. The table shows a significant relative risk of 1.02 for PM$_{2.5}$ when using $\mathbf{W}$, but a much smaller insignificant association when using $\mathbf{W}_{E}$. As the simulation study showed that using  $\mathbf{W}_{E}$ provides better covariate effect estimates, this is likely to be the more reliable result. The composite education indicator quantifies populations with little or no education (including a standardised ratio of the number of working age people with no qualifications), and increasing this by one standard deviation leads to around a 38\%  increase in the risk of respiratory hospitalisation. The crime indicator shows that areas with higher crime rates exhibit slightly lower risks, while areas that have to travel further to access amenities (Access variable) also exhibit a slightly lower risk.

\subsubsection{Disease surveillance}
\begin{figure}
	\centering
	\begin{picture}(10,16)
	\put(-1,9){\scalebox{0.23}{\includegraphics{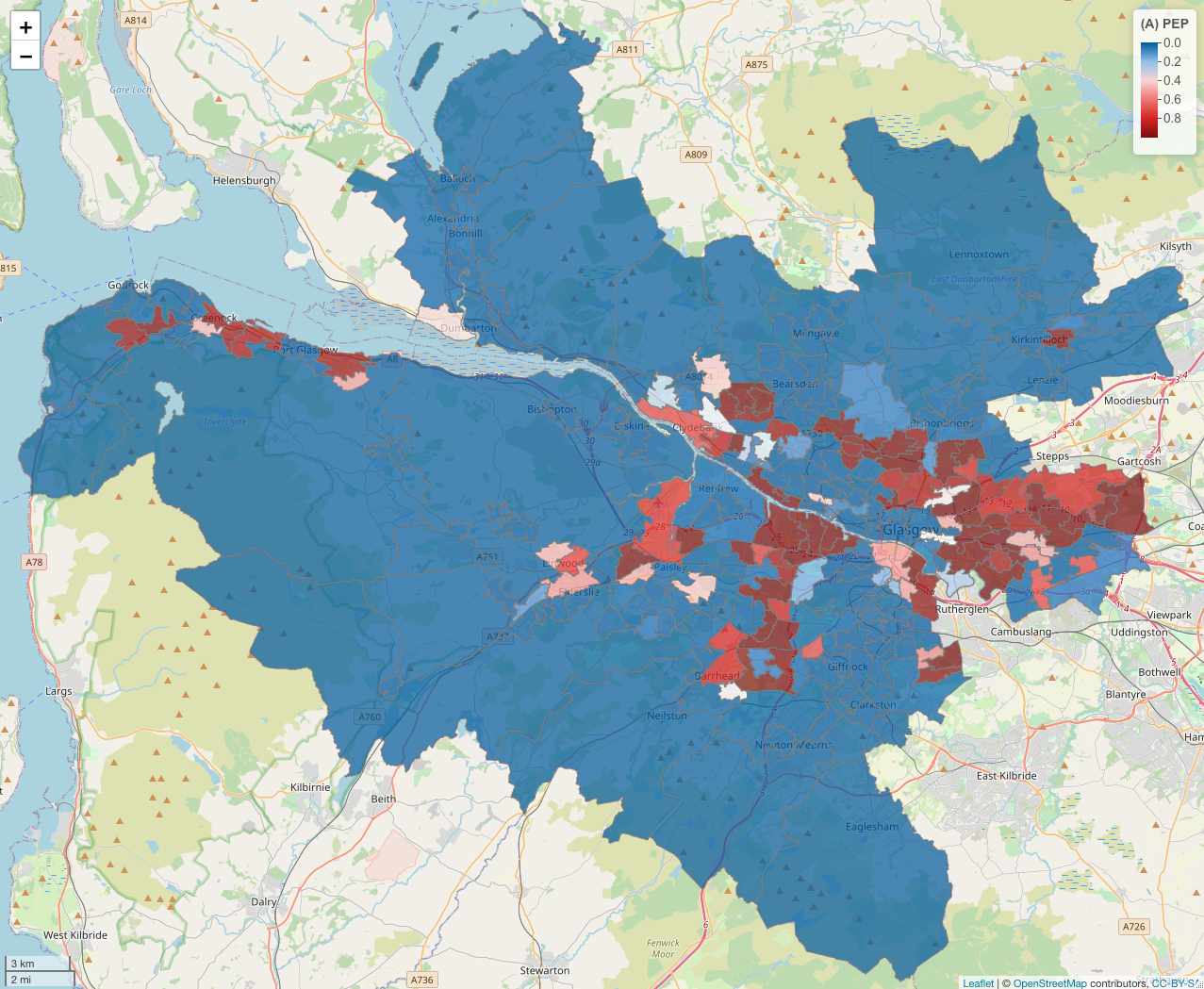}}}
	\put(-1,0){\scalebox{0.23}{\includegraphics{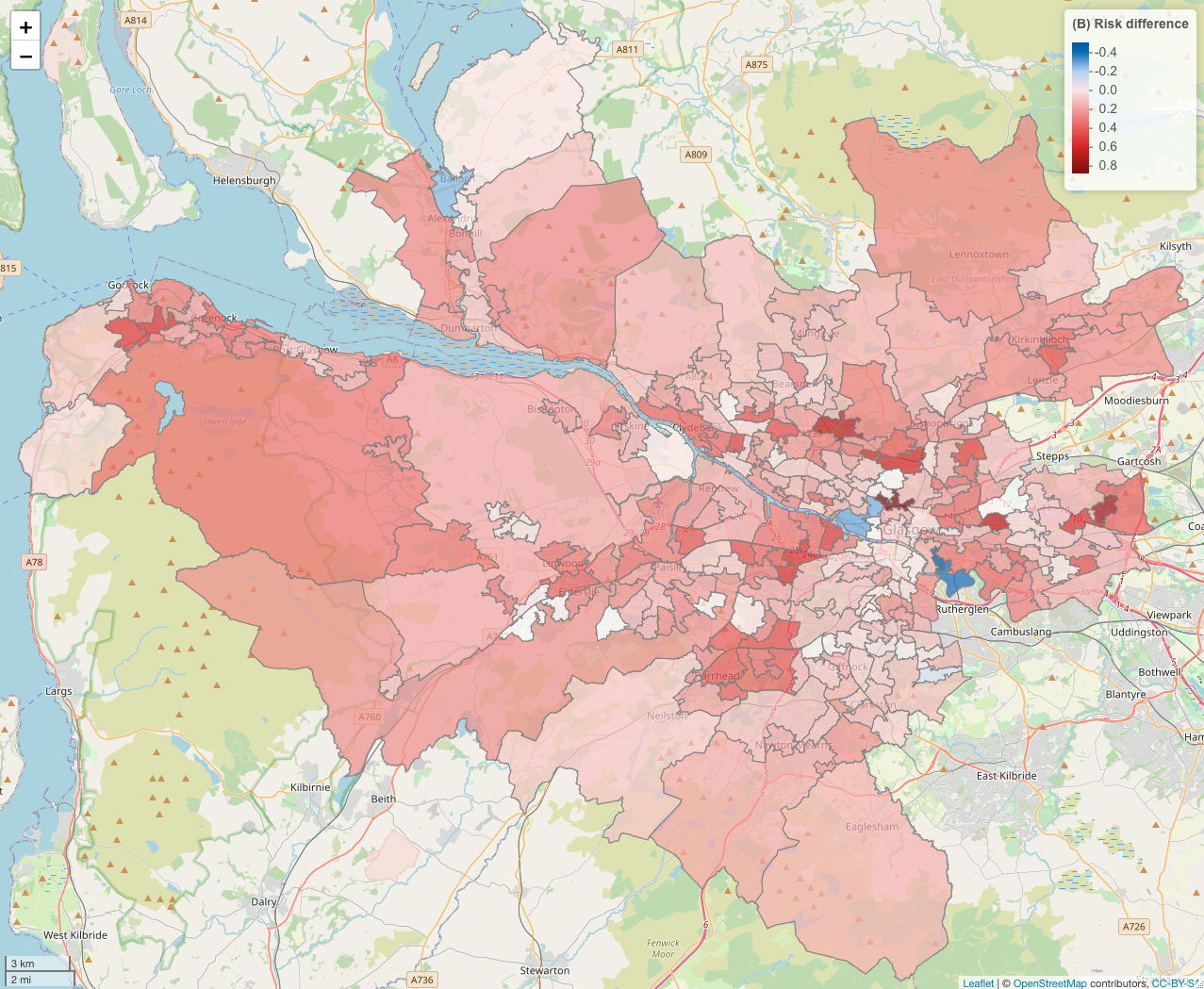}}}
	\end{picture}
	\caption{Maps displaying: (A) posterior exceedance probabilites (PEP) that the risk $\theta_{kt}$ for 2017 exceeds 1.5; and (B) the difference in risk $\theta_{kN}-\theta_{k1}$ between the last and first time period.   \label{figrisk}}
\end{figure}

Our main aim is to use the modelling for disease surveillance, and identify areas that are most in need of an intervention to improve their health. Such areas of concern exhibit elevated risks and / or an increasing risk trend, and numerous metrics have been proposed for identifying such areas (see for example  \cite{kavanagh2012}). The most popular  metrics for identifying high-risk areas are posterior exceedance probabilities (PEP) computed as $\pi_{kt} = \mathbb{P}(\theta_{kt}> C|\mathbf{Y})$, the posterior probability that the risks $\{\theta_{kt}\}$ exceed a certain threshold risk level $C$.  The specification of $C$ is somewhat arbitrary and chosen following discussions with public health experts, and here we choose $C=1.5$ which represents a 50\% elevated risk compared to the Scottish average. The PEP for 2017 is displayed in the top panel of Figure \ref{figrisk}, which shows that most IZs exhibit either a very high (dark red) or a very low (dark blue) probability of exceeding this threshold risk level. The map highlights two types of exceedances, clusters of geographically adjacent IZs exhibiting elevated risks, and individual IZs that have much higher risks than their neighbours. The east end of Glasgow in the east of the health board is the largest and most well known high risk cluster, and is in part caused by a cycle of multi-generational poverty (\citealp{NHS2016}). In contrast, the single IZ in the north east of the health board near Kirkintilloch exhibits an elevated risk unlike its geographical neighbours, and would warrant further investigation by the health board into why it exhibits a very high PEP.

The other area of concern for the health board is IZs that exhibit increasing risk trends, and panel (B) of Figure \ref{figrisk} displays the temporal changes in the posterior median risks, $\{\theta_{kN}-\theta_{k1}\}$ for each IZ $k$. The figure highlights that most IZs exhibit some level of increase in respiratory hospitalisation risk over the 7-year study period, which agrees with the exploratory analysis of the SMR in Figure \ref{figSMR}. However, a few areas exhibit decreasing risk trends such as Dalmarnock in the east of the health board (just above Rutherglen on the map), which in this case is due to the regeneration of the area following its use as the athletes village in the 2014 Glasgow Commonwealth games.

\subsubsection{Health inequalities}
Health inequalities measure the difference in disease
risk between population sub-groups, and the World Health Organisation (\citealp{WHO2013}) define \emph{total inequality} as the overall variation in disease risk, and \emph{social inequality} as the variation in risk between
different social groups. Here we quantify the size of these inequalities and how they are changing over time. We do this for total inequality by presenting the standard deviation, interquartile range and range in the estimated risk surfaces $\{\theta_{kt}\}$ separately for each year in Table \ref{tabinequality}.  The table shows substantial and sustained inequalities in risk across the GGCHB for all years, with the standard deviations in risk ranging between 0.36 and 0.45. Furthermore, the risks in 2017 range between 0.57 and 2.60, which correspond to a 75\% decreased risk (as $1 / 0.57\approx 1.75$) and a 260\% increased risk compared to the Scottish national average. The level of inequality rises between 2011 and 2013 by around 14\% (SD), 19\% (IQR) and 49\% (Range) depending on the variation measure used, but exhibits a relatively steady state since then under all 3 measures. 

Social inequality is also substantial across the GGCHB, which we quantify by computing the average risk in each decile of the education domain of the SIMD, the most important socio-economic deprivation predictor in our model. The average risk monotonically increases with increasing decile of the education covariate (higher deciles denote less educated populations) for all years, and the average risks in the first, fifth and tenth decile are displayed in Table \ref{tabinequality}. The table illustrates that the total inequalities described above are almost completely driven by socio-economic deprivation, because the lowest decile (most highly educated) has average risks ranging between 0.61 and 0.66, where as the highest decile (least highly educated) has average risks ranging between 1.86 and 2.08. These social inequalities have changed little over the 7-year study period, with almost no change in the mean risks in the 1$st$, 5$th$ and 10$th$ deciles of the education covariate over time.

\begin{table}
\centering
\caption{Summary of the total inequality and social inequality in disease risk $\{\theta_{kt}\}$ by year. Here SD - standard deviation and IQR - interquartile range. The social inequality figures represent the average risks in the 1$st$, 5$th$ and 10$th$ deciles of the education covariate.\label{tabinequality}}
\begin{tabular}{lrrrr}\hline
\multirow{2}{*}{\textbf{Year}} &  \multicolumn{3}{c}{\textbf{Total inequality}} &\textbf{Social inequality} \\
& \textbf{SD}& \textbf{IQR} & \textbf{Range} & \textbf{1st - 5th - 10th deciles}\\\hline
2011 &0.36 &0.54 &1.42 &0.61 - 1.14 - 1.86 \\
2012 &0.39 &0.57 &1.76 &0.62 - 1.12 - 2.03 \\
2013 &0.41 &0.64 &2.12 &0.64 - 1.13 - 2.03 \\
2014 &0.42 &0.65 &1.90 &0.64 - 1.12 - 1.99 \\
2015 &0.41 &0.64 &1.81 &0.67 - 1.12 - 2.03 \\
2016 &0.45 &0.64 &1.93 &0.65 - 1.12 - 2.08\\
2017 &0.42 &0.65 &2.03 &0.66 - 1.13 - 2.00 \\\hline
\end{tabular}
\end{table}

\section{Discussion}
This paper has presented a novel graph-based optimisation algorithm for estimating the neighbourhood matrix when modelling spatio-temporal areal unit count data, and has provided software to allow others to utilise our methods. Our approach thus specifies an appropriate spatial correlation structure for the data at hand via $\mathbf{W}_E$,  rather than naively specifying $\mathbf{W}$ using a simple geographical approach such as border sharing. The simulation study showed conclusive evidence that our proposed approach of using $\mathbf{W}_E$ rather than $\mathbf{W}$ delivers improved inference in terms of both risk (rate) and covariate effect estimation  for spatio-temporal data when the number of time periods is at least $N=5$.  Our approach estimates the residual spatial autocorrelation structure in the data using the residuals from a covariate only model, which is akin to applying variogram analysis to detrended geostatistical data to identify an appropriate spatial correlation model. Thus we recommend that, as in geostatistics, standard practice in spatio-temporal areal unit modelling should involve estimating both the mean model and the residual spatial dependence structure, rather than specifying the latter using a convenient rule such as border sharing with little assessment of its suitability which is currently the norm in the field (e.g. \citealp{quick2017}, \citealp{lee2019}).

The superiority of our approach was comprehensively illustrated for spatio-temporal data with and without step changes in the residual surface, although unsurprisingly the biggest improvements occur when such step changes are present.  In contrast, our approach does not work well for purely spatial data ($N=1$),  because the residual spatial surface $\tilde{\bd{\phi}}$ is not well estimated due to the random noise in the residuals (\ref{resids}) that stems from $\{Y_{kt}\}$. However, as $N$ increases this random noise is reduced by averaging the residuals over time, leading to improved performance. Thus to apply this approach to purely spatial data we suggest estimating $\tilde{\bd{\phi}}$  from  multiple sets of external data that have a similar residual spatial structure to the study data. Possible candidates in this regard are the same data but for earlier time periods, or data with a related response variable such as a different disease with a similar etiology. 

Our motivating case study has illustrated the importance of obtaining improved estimation and uncertainty quantification of disease risk, because it will lead to improved accuracy of surveillance metrics such as PEPs that depend on the full posterior distribution. Our case study also illustrates that substantial and sustained inequalities in population-level disease risk remain in the GGCHB, despite extensive  governmental focus in recent years on this key public health issue (e.g. \citealp{NHS2016}). 

There is a wealth of future research directions for extending this work, the most obvious of which is to extend the class of data and models that our graph-based optimisation approach can be used with. These include extending the methods away from count data to deal with Gaussian and binomial type responses, considering multivariate rather than spatio-temporal data structures, and using different spatio-temporal random effects structures to that considered here. Additionally, our motivating study has shown that similar levels of disease risk are more commonly observed between areas with similar levels of socio-economic deprivation rather than those that happen to be geographically close. This suggests that one might want to additionally allow for correlation between areas with similar levels of socio-economic deprivation, perhaps via the introduction of a second neighbourhood matrix based on socio-economic rather than physical adjacency. This results in the data having a correlation structure based on a multilayer  graph, and our optimisation approach would need to be extended to allow for this multilayer scenario. 

Finally, there is significant scope to improve the performance of the graph-based optimisation algorithm used to estimate $\mathbf{W}_{E}$, as the current implementation makes use of a local search method that is not guaranteed to find the best possible matrix $\mathbf{W}_{E}$ with respect to the objective function.  The fact that the optimisation problem is NP-hard in general means that we are very unlikely to find an algorithm that is guaranteed to perform the optimsation exactly within a reasonable length of time for all possible inputs.  Nevertheless, it may be possible to obtain an efficient approximation algorithm that achieves a guaranteed performance ratio (for example, computing a matrix for which the objective function is at most $5\%$ worse than the best possible), or parameterised algorithms which have exponential running-time in the worst case but are guaranteed to perform much faster on inputs with specific structural properties. Further work is needed to establish the feasibility or otherwise of both approaches.

\section*{Appendix}

Algorithm \ref{alg:local-search} below summarises our graph-based optimisation algorithm.

\begin{algorithm}
%\LinesNumbered
\SetAlgoNoEnd
\SetAlgoNoLine
$H \leftarrow G$\;
\texttt{Oldscore} $\leftarrow \infty$\;
\texttt{Newscore} $\leftarrow f(H_0,\tilde{\bd{\phi}})$\;
\While {\textup{\texttt{Oldscore}} $<$ \textup{\texttt{Newscore}}} {
	\texttt{OldH} $\leftarrow H$\;
	\texttt{Oldscore} $\leftarrow$ \texttt{Newscore}\;
	$W \leftarrow \{v \in V \colon \deg_H(v) > 1\}$\;
	\For {$v \in w$} {
		\For {$u \in W \cap N_H(v)$} {
			\texttt{Best\_v\_with\_u} $\leftarrow \max_{\substack{N^- \subseteq N_H(v) \cap W \\ u \notin N^-}}\{\adjcont(v,H \setminus \{vw: w \in N^-\},\tilde{\bf{\phi}})\}$\;
			\texttt{Best\_v\_without\_u} $\leftarrow \max_{\substack{N^- \subseteq N_H(v) \cap W \\ u \in N^-}}\{\adjcont(v,H \setminus \{vw: w \in N^-\},\tilde{\bf{\phi}})\}$\;
			\texttt{Best\_u\_with\_v} $\leftarrow \max_{\substack{N^- \subseteq N_H(u) \cap W \\ v \notin N^-}}\{\adjcont(u,H \setminus \{uw: w \in N^-\},\tilde{\bf{\phi}})\}$\;
			\texttt{Best\_u\_without\_v} $\leftarrow \max_{\substack{N^- \subseteq N_H(u) \cap W \\ v \in N^-}}\{\adjcont(u,H \setminus \{uw: w \in N^-\},\tilde{\bf{\phi}})\}$\;
			\If {(\textup{\texttt{Best\_v\_with\_u}} $+$ \textup{\texttt{Best\_u\_with\_v}}) $<$ (\textup{\texttt{Best\_v\_without\_u}} $+$ \textup{\texttt{Best\_u\_without\_v}})} {
				\If {$\min\{\deg_H(v),\deg_H(u)\} > 1$} {
					$H \leftarrow H \setminus \{uv\}$\;
					\If {$\deg_H(u) = 1$} {
						$W \leftarrow W \setminus \{u\}$\;
					}
				}
			}
		}
	}
	\texttt{Newscore} $\leftarrow f(H,\tilde{\bf{\phi}})$\;
}
\Return {\texttt{OldH}}
\caption{Local search procedure which takes as input $\tilde{\bf{\phi}}$ and the graph $G$ corresponding to the original matrix $\mathbf{W}$, and iteratively improves the graph with respect to the objective function.}
\label{alg:local-search}
\end{algorithm}

\bigskip
\begin{center}
{\large\bf SUPPLEMENTARY MATERIAL}
\end{center}

\begin{description}
\item[Supplementary file 1:] Additional results from the simulation study (.pdf)
\end{description}

\bibliographystyle{chicago}
\bibliography{Lee}
\end{document}

\def\spacingset#1{\renewcommand{\baselinestretch}%
{#1}\small\normalsize} \spacingset{1}

%%%%%%%%%%%%%%%%%%%%%%%%%%%%%%%%%%%%%%%%%%%%%%%%%%%%%%%%%%%%%%%%%%%%%%%%%%%%%%

\if1\blind
{
  \title{\bf Supplementary material for  `Improved inference for areal unit count data using graph-based optimisation'}
  \author{Duncan Lee\\
    School of Mathematics and Statistics, University of Glasgow\\
    and \\
    Kitty Meeks\thanks{
    Both authors gratefully acknowledge funding from the Engineering and Physical Sciences Research Council (ESPRC) grant number EP/T004878/1 for this work, while the work of the second author was also funded by a Royal Society of Edinburgh Personal Research Fellowship (funded by the Scottish Government). The respiratory hospitalisation data were provided by Public Health Scotland. }\hspace{.2cm} \\
    School of Computing Science, University of Glasgow}
  \maketitle
} \fi

\if0\blind
{
  \bigskip
  \bigskip
  \bigskip
  \begin{center}
    {\LARGE\bf Improving inference for areal unit count data using graph-based optimisation}
\end{center}
  \medskip
} \fi

\vfill

\newpage
\spacingset{1.45} % DON'T change the spacing!
\section*{Introduction}
This supplementary material contains the following sections. Section 1 presents example realisations of the random effects surfaces generated in the simulation study, while Section 2 presents additional simulation results relating to the estimation of covariate effects. 

\section{Example realisations of the simulated random effects surfaces}
\begin{figure}
	\centering
	\begin{picture}(14,18)
	\put(-2,8){\scalebox{0.18}{\includegraphics{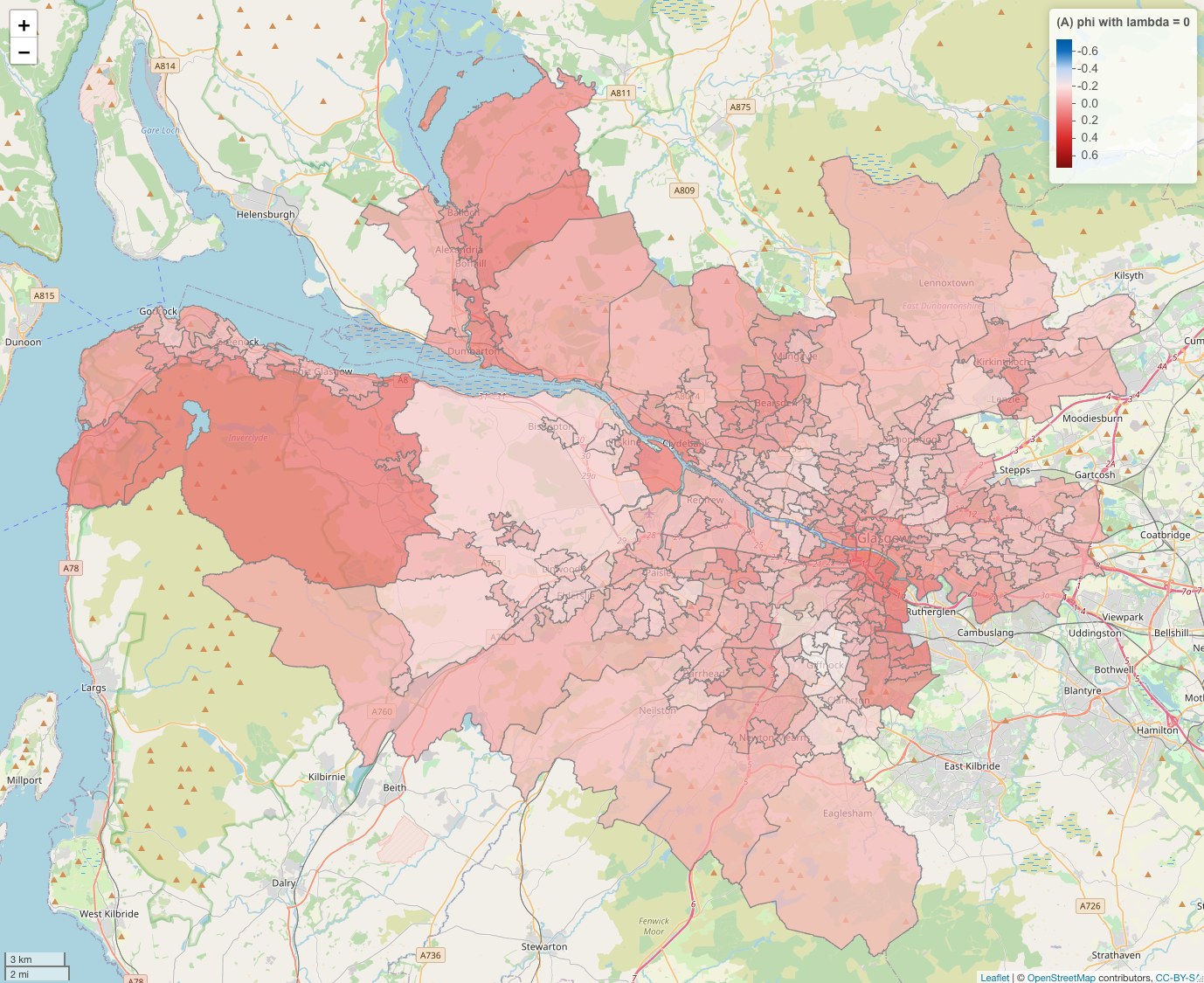}}}
	\put(7,8){\scalebox{0.18}{\includegraphics{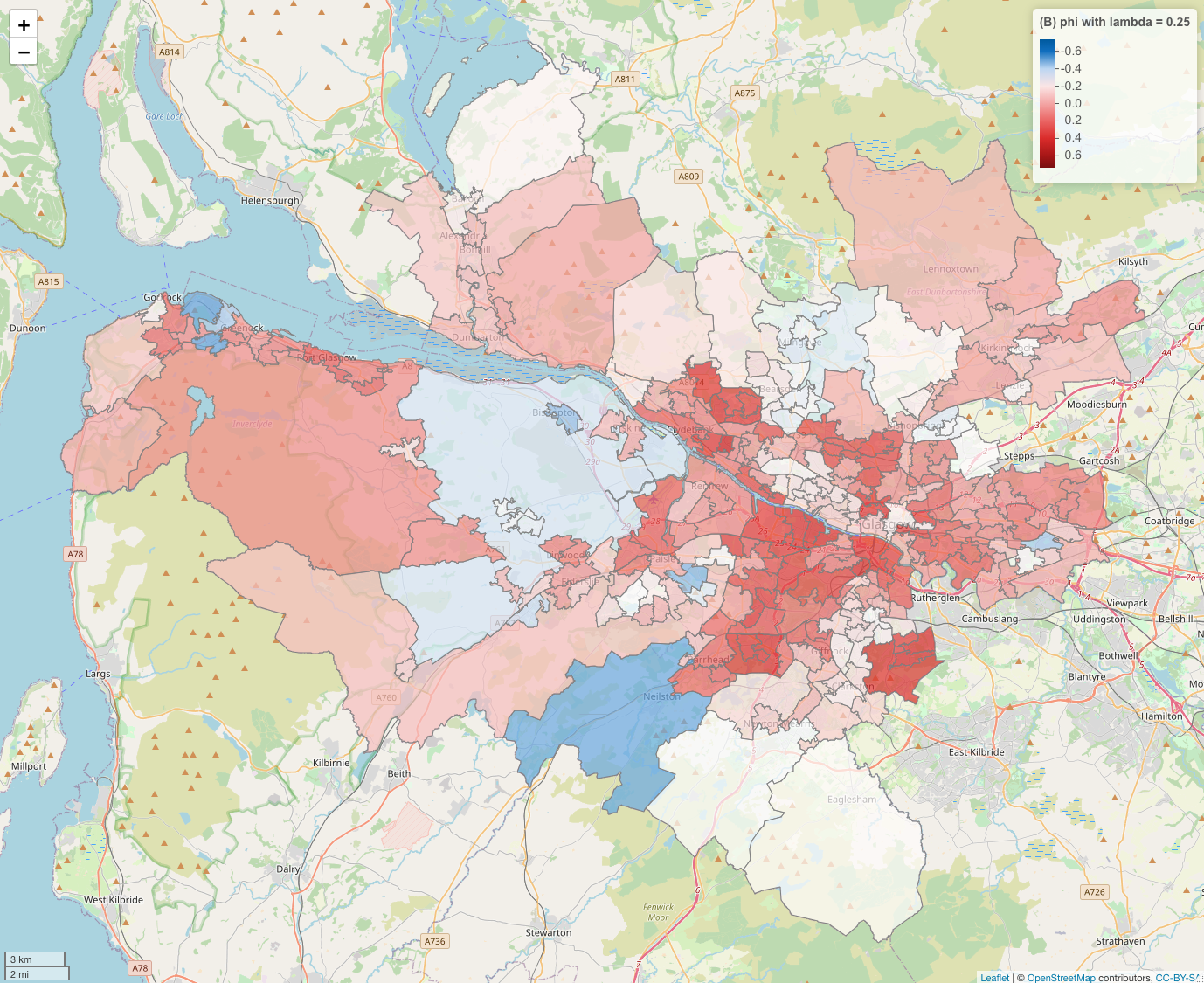}}}
	\put(3,0){\scalebox{0.18}{\includegraphics{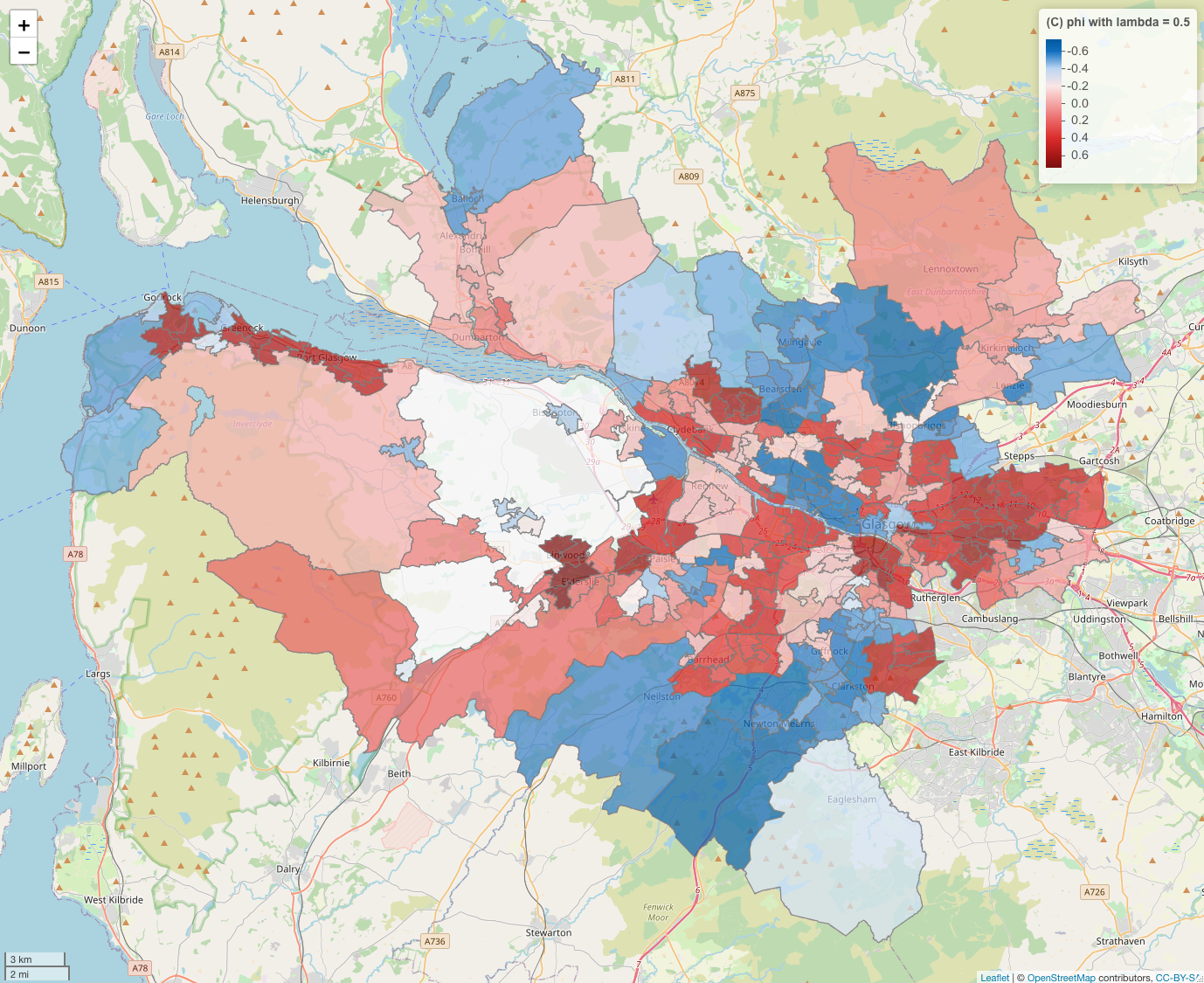}}}
	\end{picture}
	\caption{Maps displaying example realisations of the random effects $\bd{\phi}_t$ for: (A) $\lambda=0$; (B) $\lambda=0.25$; and (C) $\lambda=0.5$.}   \label{figphi}
\end{figure}

Example realisations of $\bd{\phi}_t$ for all 3 values of $\lambda$ (which controls the size of the step changes) are presented in Figure \ref{figphi}, to illustrate the spatial pattern in, and the magnitude of, these step changes. The figure shows that when $\lambda=0$ the residual spatial surface appears visually smooth, whilst when $\lambda=0.5$ large step changes are apparent. Obviously, setting $\lambda=0.25$ results in an intermediate map with some small step changes. 

\section{Additional simulation study results - covariate effects}
The simulation study results for the covariate effects are summarised in Tables \ref{simindep} (independent covariate) and \ref{simcorr}  (spatially correlated covariate), and have the same format as Table 1 in the main paper. As before root mean square errors (RMSE) are presented, along with coverage probabilities and average widths of the associated 95\% credible intervals. The tables show that covariate estimation exhibits the same 3 main findings as risk (rate) estimation presented in the main paper. Firstly, if one has spatial rather than spatio-temporal data, ($N=1$) then using the estimated $\mathbf{W}_{E}$ leads to worse covariate effect estimation than using a simpler geographical specification based on border sharing. This is true for both independent and spatially correlated covariates, with both exhibiting higher RMSE values and poorer coverage when using $\mathbf{W}_{E}$. However, if one has spatio-temporal data ($N>1$) with step changes ($\lambda>0$), then using $\mathbf{W}_{E}$ results in lower RMSE and narrower credible intervals with more appropriate coverage probabilities (closer to 95\%) compared to using $\mathbf{W}$, making it the method of choice in such situations. In the absence of step changes ($\lambda=0$) the two approaches perform similarly, with similar RMSE values for both covariate types and slightly better coverage probabilities when using $\mathbf{W}_{E}$ to estimate the effects of spatially correlated covariates.

\begin{table}
\caption{Accuracy of the independent covariate effects $\beta_1$ from the model with the border sharing ($\mathbf{W}$) and estimated ($\mathbf{W}_{E}$) neighbourhood matrices.\label{simindep}}
\begin{tabular}{llrrrr}\hline
\textbf{Step}&\textbf{Time}&\multicolumn{4}{c}{\textbf{Disease prevalence}}\\
\multicolumn{1}{l}{\textbf{change}}&\multicolumn{1}{l}{\textbf{period}}& \multicolumn{2}{c}{$\mathbf{e_i\in[10, 30]}$}&\multicolumn{2}{c}{$\mathbf{e_i\in[150, 250]}$}\\\hline
\textbf{RMSE}&&$\mathbf{W}$ &$\mathbf{W}_{E}$ & $\mathbf{W}$ &$\mathbf{W}_{E}$\\
&$N=1$ &55.1 &54.5 &22.1 &26.5\\
$\lambda=0$&$N=5$ &25.7 &25.9 &11.6 &11.7\\
&$N=9$ &21.5 &21.5 &9.3 & 8.7\\\hline
&$N=1$ &69.8 &70.2 &41.2 &45.9\\
$\lambda=0.25$&$N=5$ &33.6 &30.7 &21.2 &14.1\\
&$N=9$ &25.3 &20.8 &15.3 &8.1\\\hline
&$N=1$ &113.4 &117.4 &97.7 &79.0\\
$\lambda=0.5$&$N=5$ &41.8 &35.2 &33.3 &15.0\\
&$N=9$ &37.9 &21.5 &26.6 &9.4\\\hline
\textbf{Coverage}&&$\mathbf{W}$ &$\mathbf{W}_{E}$ & $\mathbf{W}$ &$\mathbf{W}_{E}$\\
\textbf{(width)}&$N=1$ &100 (0.113) &100 (0.126) &97 (0.047) &91 (0.045)\\
$\lambda=0$&$N=5$ &96 (0.052) &96 (0.054) &95 (0.023) & 94 (0.021)\\
&$N=9$ &93 (0.039) &95 (0.040) &94 (0.017) &95 (0.016)\\\hline
&$N=1$ &98 (0.144) &96 (0.138) &98 (0.092) &78 (0.058)\\
$\lambda=0.25$&$N=5$ &94 (0.065) &96 (0.061) &98 (0.042) &95 (0.026)\\
&$N=9$ &93 (0.049) &98 (0.045) &95 (0.031) &97 (0.020)\\\hline
&$N=1$ &94 (0.201) &82 (0.157) &90 (0.166) &64 (0.070)\\
$\lambda=0.5$&$N=5$ &96 (0.091) &95 (0.068) &99 (0.074) &95 (0.028)\\
&$N=9$ &90 (0.067) &96 (0.049) &94 (0.056) &98 (0.021) \\\hline
\end{tabular}
\end{table}

\begin{table}
\caption{Accuracy of the spatially correlated covariate effects $\beta_2$ from the model with the border sharing ($\mathbf{W}$) and estimated ($\mathbf{W}_{E}$) neighbourhood matrices.\label{simcorr}}
\begin{tabular}{llrrrr}\hline
\textbf{Step}&\textbf{Time}&\multicolumn{4}{c}{\textbf{Disease prevalence}}\\
\multicolumn{1}{l}{\textbf{change}}&\multicolumn{1}{l}{\textbf{periods}}& \multicolumn{2}{c}{$\mathbf{e_i\in[10, 30]}$}&\multicolumn{2}{c}{$\mathbf{e_i\in[150, 250]}$}\\\hline
\textbf{RMSE}&&$\mathbf{W}$ &$\mathbf{W}_{E}$ & $\mathbf{W}$ &$\mathbf{W}_{E}$\\
&$N=1$ &80.8 &75.0 &45.9 &57.2\\
$\lambda=0$&$N=5$ &53.6 &50.6 &25.7 &21.6\\
&$N=9$ &36.2 & 33.7 &22.1 &19.2\\\hline
&$N=1$ &120.2 &124.5 &90.7 &92.4\\
$\lambda=0.25$&$N=5$ &53.6 &51.0 &47.1 &29.4\\
&$N=9$ &49.0 &41.4  &32.4 &21.7\\\hline
&$N=1$& 180.7 &163.9 &147.6 &134.1\\
$\lambda=0.5$&$N=5$ &82.4 &54.5 &90.7 &36.4\\
&$N=9$ &64.1 & 44.1 &61.5 &26.9 \\\hline
\textbf{Coverage}&&$\mathbf{W}$ &$\mathbf{W}_{E}$ & $\mathbf{W}$ &$\mathbf{W}_{E}$\\
\textbf{(width)}&$N=1$ &84 (0.121) & 96 (0.166) &87 (0.071) & 76 (0.067)\\
$\lambda=0$&$N=5$ &76 (0.067) & 81 (0.070) &87 (0.039) &92 (0.036)\\
&$N=9$ &82 (0.050) &88 (0.051) &85 (0.030) &86 (0.028)\\\hline
&$N=1$ &91 (0.203) &87 (0.202) &91 (0.152) &71 (0.097)\\
$\lambda=0.25$&$N=5$ &94 (0.100) &96 (0.096) &90 (0.075) &91 (0.049)\\
&$N=9$ &91 (0.075) &94 (0.071) &94 (0.057) &90 (0.037)\\\hline
&$N=1$ &90 (0.313) & 83 (0.250) &91 (0.266) &67 (0.118)\\
$\lambda=0.5$&$N=5$ &95 (0.152) &97 (0.118) &89 (0.131) &90 (0.055)\\
&$N=9$ &95 (0.116) &96 (0.087) &88 (0.099) &86 (0.041)\\\hline
\end{tabular}
\end{table}